\documentclass[aip,pof,jmp,onecolumn]{revtex4-1}

\usepackage{graphicx}
\graphicspath{ {Figures/} }
\usepackage{natbib}
\usepackage{amsmath,amssymb}
\usepackage{subfig}
\usepackage{setspace}

\usepackage[font=small,font=singlespacing,justification=raggedright]{caption}

\usepackage{dcolumn}% Align table columns on decimal point
\usepackage{hyperref}% add hypertext capabilities
\usepackage{ifthen}
\usepackage[usenames, dvipsnames]{color}

\providecommand{\norm}[1]{\lVert#1\rVert}
\newboolean{showcomments}
\setboolean{showcomments}{true}

\newcommand{\dennice}[1]{  \ifthenelse{\boolean{showcomments}}
{\textcolor{blue}{(Dennice says:  #1)}}{}}
\newcommand{\brian}[1]{\ifthenelse{\boolean{showcomments}}
{\textcolor{Red}{(Brian says: #1)}}{}}
\newcommand{\petros}[1]{\ifthenelse{\boolean{showcomments}}
{\textcolor{Green}{Petros says: #1)}}{}}
\newcommand{\vaughan}[1]{\ifthenelse{\boolean{showcomments}}
{\textcolor{Purple}{Vaughan says: #1)}}{}}
\newcommand{\tc}{\textcolor}

\begin{document}

\title{Self-sustaining turbulence in a restricted nonlinear model of plane Couette flow}% Force line breaks with \\
%\title{Self-sustaining Turbulence in the Restricted Nonlinear Systems}% Force line breaks with \\

\author{Vaughan L. Thomas}
\affiliation{Department of Mechanical Engineering, Johns Hopkins University}
\author{Binh  K. Lieu}
\affiliation{Department of Electrical and Computer Engineering, University of Minnesota}
\author{Mihailo R. Jovanovi\'{c}}
\affiliation{Department of Electrical and Computer Engineering, University of Minnesota}
\author{Brian Farrell}
\affiliation{School of Engineering and Applied Science, Harvard University}\author{Petros Ioannou}
\affiliation{Department of Physics, National and Kapodistrian University of Athens}
\author{Dennice F. Gayme}
\affiliation{Department of Mechanical Engineering, Johns Hopkins University}
\date{\today}% It is always \today, today,

\begin{abstract}

This paper demonstrates the maintenance of  self-sustaining turbulence in a restricted nonlinear (RNL) model of plane Couette flow. The RNL system is derived directly from the Navier Stokes equations and permits computationally tractable studies of the dynamical system obtained using stochastic structural stability theory (S3T),
which is a second order approximation of the statistical state dynamics of the flow. The RNL model shares the dynamical restrictions of the S3T model but can be easily implemented through reducing a DNS code to the equations governing the RNL system.  Comparisons of turbulence arising from DNS and RNL simulations demonstrate that the RNL system supports self-sustaining turbulence with a mean flow as well as structural and dynamical features that are consistent with DNS. These results demonstrate that the simplified RNL/S3T system captures fundamental aspects of fully developed turbulence in wall-bounded shear flows and motivates use  of the RNL/S3T system for further study of  wall turbulence.
%\dennice{what was written here is not clear at all and is unnecessarily complicated for an abstract.  I have changed it back to what we previously agreed on which is correct and a lot easier to understand}.
\iffalse 
When S3T is implemented as a coupled set of equations for the streamwise mean flow and perturbations about that mean, the nonlinearity in the dynamics is restricted to interaction between the streamwise mean and associated perturbations. 
When S3T is implemented using a streamwise mean and perturbation decomposition, the perturbation state is represented by its ensemble mean covariance.
 \fi

\end{abstract}

\maketitle

%% SECTION NAMES:
%% Should be lower-case except for proper nouns and abbreviations
%% Should not end with a period
%=========================================================================
%=========================================================================
\section{Introduction}

The Navier Stokes (NS) equations provide a comprehensive model for the dynamics of turbulence. Unfortunately, these equations are analytically intractable. They have, however, been extensively studied computationally since the pioneering work of \citet{Kim-etal-1987} and a number of highly resolved numerical simulations exist, see e.g. \cite{Simens-2009,DelAlamo-etal-2004,Tsukahara-etal-2006,Hoyas-Jimenez-2008}. Ever increasing computing power will continue progress toward simulating an increasingly wider range of turbulent flows. However, a complete understanding of the physical mechanisms underlying turbulence in the NS equations, even in simple parallel flow configurations, remains elusive. Thus, considerable effort has been devoted to the search for more tractable models that characterize the dynamics of turbulence.

The Linearized Navier Stokes (LNS) equations are a particularly appealing model because they can  be analyzed using well developed tools from linear systems theory \cite{Farrell-Ioannou-1996a,Farrell-Ioannou-1996b}. They have been used extensively to characterize energy growth and disturbance amplification in wall bounded shear flows, in particular the large disturbance amplification that arises from the non-normal linear operators governing these flows \cite{Farrell-1988a,Gustavsson-1991,Trefethen-etal-1993,Reddy-Henningson-1993,Farrell-Ioannou-1993e,Bamieh-Dahleh-2001,Jovanovic-Bamieh-2005}. The LNS equations capture the energy production of the full nonlinear system \cite{Henningson-1996b} and linear non-normal growth mechanisms have been shown to be necessary for subcritical transition to turbulence~\cite{Henningson-Reddy-1994}. \iffalse In connection with the problem of understanding how turbulence is maintained, \fi The LNS equations also provide insight into the mechanism maintaining turbulence. In particular, the linear coupling between the Orr-Sommerfeld and Squire equations is required to generate the wall layer streaks that are a necessary component of the process maintaining turbulence in wall-bounded shear flows\cite{Butler-Farrell-1993,Kim-Lim-2000}. In this context, the term ``streak'' describes the ``well-defined elongated region of spanwise alternating bands of low and high speed fluid'' \citep{WKH91}. The LNS equations have also been used  to predict second-order statistics~\cite{jovbamCDC01,jovgeoAPS10} and the spectra of turbulent channel flows \cite{Butler-Farrell-1993,Farrell-Ioannou-1998a,DelAlamo-Jimenez-2006,Hwang-Cossu-2010,Cossu-etal-2009,moajovJFM12}. The above results and a host of others illustrate the power of the LNS equations as a model for studying physical mechanisms in wall-bounded turbulence. While the LNS equations provide insight into the mechanisms underlying turbulence, there are two fundamental aspects of turbulence that the LNS system is  unable to model: the turbulent mean velocity profile and the mechanism that maintains turbulence.

Empirical models have also proven useful in capturing certain aspects of turbulent flows. For example, Proper Orthogonal Decomposition (POD) has been used to construct low dimensional ordinary differential equation models of turbulent flows, see e.g. \cite{Lumley-1967,Smith-etal-2005}. However, empirical models of this type are based on data resulting from experiments or simulations rather than proceeding directly from the NS equations. This limitation is shared by eddy viscosity models. 

Researchers have also sought insight into turbulence through examining numerically obtained three-dimensional equilibria and periodic orbits of the NS equations, see e.g. \cite{Jimenez-etal-2005,Kawahara-etal-2012}. For plane Couette flow, the first such numerical solution was computed by \citet{Nagata-1990}. Details concerning these numerically obtained fixed points and periodic orbits for plane Couette flow can be found in \cite{Gibson-etal-2009,Kawahara-etal-2012}. These solutions reflect local properties of the attractor and the extension of these solutions to the global turbulent dynamics has yet to be completed.

The 2D/3C model~\cite{Reynolds-Kassinos-1995,Bobba-etal-2002,Gayme-etal-2010} is a recent attempt to obtain an analytically tractable simplified model derived from the NS equations. The assumptions underlying this model are based on experimental \cite{Kim-Adrian-1999,Guala-etal-2006,Hutchins_Marusic-2007} and analytical evidence \cite{Farrell-Ioannou-1993e,Jovanovic-Bamieh-2005,Cossu-etal-2009,McKeon-Sharma-2010} of the central role of streamwise coherent structures in wall turbulence.  This streamwise constant model has been used to accurately simulate the mean turbulent velocity profile \cite{Gayme-etal-2010}, to identify the large-scale spanwise spacing of the streamwise coherent structures and to
study the energetics of fully developed turbulent plane Couette flows \cite{Gayme-etal-2011}. The primary limitation of the 2D/3C model is that it supports only one-way interactions from the perturbation field to the mean flow and as a consequence it requires persistent stochastic excitation to sustain the turbulent state perturbations. In fact, the laminar solution of the unforced 2D/3C model has been shown to be globally asymptotically stable~\cite{BobbaThesis}\footnote{A proof of this fact and the explicit construction of a Lyapunov function based on private communications with A. Papachristodoulou and B. Bamieh is provided in \cite{GaymeThesis}.}.

The current work describes a more comprehensive model that is similar to the 2D/3C model in its use of a streamwise constant mean flow, but which also incorporates two-way interaction between this streamwise constant mean flow and the perturbation field.  This coupling is chosen to parallel that used in the Stochastic Structural Stability Theory (S3T) model~\cite{Farrell-Ioannou-2003-structural}. The S3T equations comprise the joint evolution of the streamwise constant mean flow (first cumulant) and  the ensemble  second order perturbation statistics (second  cumulant), and can be viewed as a second order closure of the dynamics of the statistical state.  These equations are closed either by parameterizing the higher cumulants as a stochastic excitation \cite{Farrell-Ioannou-1993e,DelSole-Farrell-1996,DelSole-04} or by setting the third cumulant to zero, see e.g \cite{Marston-etal-2008,Tobias-etal-2011,Srinivasan-Young-2012}. This restriction of the NS equations to the first two cumulants  involves parameterizing or neglecting the perturbation-perturbation  interactions in the full nonlinear system and retaining only the interaction between the perturbations and the instantaneous mean flow. This closure results in a nonlinear autonomous dynamical system that governs evolution of the statistical state of the turbulence comprised of this mean flow and the second order perturbation statistics. A simulation of the restricted nonlinear (RNL) system may be regarded as a statistical state dynamics obtained from a single
member of the ensemble making up the S3T dynamics.\iffalse in which the second order cumulant is approximated using a two point correlation of the flow field.\fi 

The S3T model has recently been used to study the dynamics of fully developed wall turbulence~\cite{Farrell-etal-2012-ctr}, in particular that of the roll and streak structures. These prominent features of wall-turbulence were first identified in the buffer layer~\cite{Kline-et-al-1967}. \iffalse and subsequently studied extensively both observationally and using simulation \cite{??} . \fi   Rolls and streaks have often been suggested to play a central role in maintaining wall turbulence but neither the laminar nor the turbulent streamwise mean velocity profile give rise to these structures as a fast inflectional instability of the type generally associated with rapid transfer of energy from the mean flow to sustain the perturbation field.  However, these structures are associated with transient growth in wall-bounded shear flows, which leads to robust transfer of energy from the mean wall-normal shear to the perturbation field. In particular, 
this transfer occurs as the roll circulation drives the streak perturbation through the lift-up mechanism~\cite{Landahl-1980}.  The conundrum posed by the linear stability of the roll and streak structures and their recurrence  in turbulence was first posited as being a result of their participation in  a regeneration cycle in which the roll is maintained by  perturbations resulting from the break-up of the streaks \cite{Swearingen-Blackwelder-1987,Blakewell-Lumley-1967}. This proposed cycle is a nonlinear instability process sustained by energy transfer due to the linear non-normal lift-up growth processes. 

The regeneration cycle of rolls and streaks has been attributed to a variety of mechanisms collectively  referred as the self-sustaining processes (SSP).  One class of SSP mechanisms attributes the perturbations sustaining the roll circulation to an inflectional instability of the streak~\cite{Waleffe-1995a,Hamilton-etal-1995,Waleffe-1997,Hall-Sherwin-2010}. Other researchers subsequently observed that most streaks in the buffer layer are too weak to be unstable to the inflectional mechanism and postulated that a transient growth mechanism is an equally plausible explanation for the origin of roll-maintaining perturbations \cite{Schoppa-Hussain-2002}. Moreover, transiently growing perturbations have the advantage of potentially tapping the energy of the wall-normal mean shear.  In fact, the mostly rapidly growing perturbations in shear flow are oblique waves with this property of drawing on the mean shear \cite{Farrell-Ioannou-1993a} and consistently, oblique waves are commonly observed to accompany streaks in wall-turbulence \cite{Schoppa-Hussain-2002}.  The mechanism in which transiently growing perturbations that draw on the mean shear maintain the roll/streak complex through an SSP requires an explicit explanation for the collocation of the perturbations with the streak, a question that the linearly unstable streak based SSP circumvents. 

In the SSP identified in the S3T system the roll is maintained by transiently growing perturbations that tap the energy of the mean shear rather than by an inflectional instability of the streak.  The crucial departure from previously proposed transient  growth mechanisms is that these transiently growing perturbations result from parametric instability of the time-dependence streak \cite{Farrell-Ioannou-2012} rather than arising from break-down of the streak \cite{Jimenez-2013}.  This parametric SSP explains inter alia the systematic collocation of the streak with the roll-forming perturbations and the systematic transfer of energy from the wall-normal shear to maintain the streak.

In this paper, we verify that the RNL system supports self-sustaining turbulence by comparing RNL simulations to DNS. We further show that the SSP supported by the RNL system is consistent with the familiar roll/streak SSP observed in wall turbulence. Because  RNL dynamics is so closely related to the S3T dynamics, any SSP that is operating in both RNL and S3T is persuasively the same and to the extent that the turbulence  seen in RNL simulations and that seen  in the DNS are also similar this argues that the parametric SSP identified in S3T/RNL is also operating in the dynamics underlying DNS.

This paper is organized as follows. The next section derives the RNL model from the NS equations and establishes its relation to the S3T system. In section \ref{sec:numerics} we describe our numerical approach and then in section \ref{sec:results} we demonstrate that the RNL system produces turbulence that is strikingly similar to that of DNS. This result verifies that the interaction between the perturbations and the streamwise constant mean flow retained in the RNL/S3T framework is sufficient for maintaining turbulence. In section \ref{sec:2d3c} we compare fully developed RNL turbulence to that arising from a stochastically forced 2D/3C model, to highlight the importance of the fundamental interactions between the perturbations and the time-dependent mean flow. These interactions, which are present in the RNL model but not in the 2D/3C model, are essential for sustaining turbulence. Finally, we conclude the paper and point to directions of future study.

\section{Methods}
\label{sec:framework}

\subsection{Modeling framework}

 Consider a plane Couette flow between walls with velocities $\pm U_w $. The streamwise direction is $x$, the wall-normal direction is $y$, and the spanwise direction is $z$. Quantities  are non-dimensionalized by the channel half-width, $\delta$, and the wall velocity, $U_w$. The non-dimensional lengths of the channel in the streamwise and spanwise  directions are respectively $ L_x$ and $L_z $. Streamwise averaged, spanwise averaged, and time averaged quantities are denoted respectively by angled brackets, $\langle\,\bullet\,\rangle = \tfrac{1}{L_x} \int_0^{ L_x} \bullet ~  {\rm{d}} x $, square brackets, $[\bullet]=\tfrac{1}{L_z} \int_0^{L_z} \bullet ~{\rm{d}} z$, and an overline $\overline{\;\bullet\;} = \tfrac{1}{T} \int_0^T \bullet~{\rm{d}} t$, with  $T$ sufficiently large. The velocity field $\mathbf{u}_T$ is decomposed into its streamwise mean, $\mathbf{U}(y,z,t)=(U,V,W)$, and the deviation from this mean (the perturbation), $\mathbf{u}(x,y,z,t)=(u,v,w)$.  The pressure gradient is similarly decomposed  into its streamwise mean, $\nabla P(y,z,t)$, and the deviation from this mean, $\nabla p(x,y,z,t) $. The corresponding Navier Stokes (NS) equations are
 \begin{subequations}
\label{eq:NSE0}
\begin{eqnarray}
&&\mathbf{U}_t+ \mathbf{U} \cdot \nabla \mathbf{U} + \nabla P - \frac{1}{R}\Delta \mathbf{U} = - \langle\mathbf{u} \cdot \nabla \mathbf{u}\rangle,
\label{eq:NSm}\\
&& \mathbf{u}_t+   \mathbf{U} \cdot \nabla \mathbf{u} +
\mathbf{u} \cdot \nabla \mathbf{U}  + \nabla p -  \frac{1}{R}\Delta  \mathbf{u}
= -  \left(\mathbf{u} \cdot \nabla \mathbf{u} - \langle\mathbf{u} \cdot \nabla \mathbf{u}\rangle \right) + \mathbf{\epsilon}
 \label{eq:NSp}\\
&&\nabla \cdot \mathbf{U} = 0,~~~\nabla \cdot \mathbf{u} = 0 , \label{eq:1c}~~~
\end{eqnarray}
\end{subequations}
 where the Reynolds number is defined as $R = {U_w \delta}/{\nu}$, with kinematic viscosity $\nu$. The parameter $\mathbf{\epsilon}$ in \eqref{eq:NSp} is an externally imposed divergence-free stochastic excitation that is used  to induce transition to turbulence.

We derive the RNL system from \eqref{eq:NSE0} by first introducing a stochastic excitation, $\mathbf{e}$, to parameterize  the  nonlinearity, $ \mathbf{u} \cdot \nabla \mathbf{u} - \langle\mathbf{u} \cdot \nabla \mathbf{u}\rangle$  as well as  divergence-free external excitation  $\epsilon$ in  \eqref{eq:NSp} to obtain:
 \begin{subequations}
\label{eqn:RNL}
\begin{eqnarray}
&&{ \mathbf{U}_t}+   \mathbf{U} \cdot \nabla \mathbf{U}
 + \nabla  P - \frac{1}{R} \Delta \mathbf{U}  = -
 \langle\mathbf{u} \cdot \nabla \mathbf{u}\rangle,
\label{eqn:RNL-mean}\\
&&{ \mathbf{u}_t}+   \mathbf{U} \cdot \nabla \mathbf{u} +
\mathbf{u} \cdot \nabla \mathbf{ U}  + \nabla  p -  \frac{1}{R} \Delta \mathbf{u} = \mathbf{e},\label{eqn:RNL-perturb}\\
 &&\nabla \cdot \mathbf{U} = 0,~~~\nabla \cdot \mathbf{u} = 0 .~~~\label{eqn:RNL-div}
 \end{eqnarray}
\end{subequations}

This results in a nonlinear system where \eqref{eqn:RNL-mean} describes the dynamics of a streamwise mean flow driven by the divergence of the streamwise averaged Reynolds stresses; we denote these streamwise averaged perturbation Reynolds stress components as e.g. $\langle{uu}\rangle$, $\langle{uv}\rangle$.
 On the other hand, equation \eqref{eqn:RNL-perturb} accounts for the interactions between the streamwise varying perturbations, $\mathbf{u}(x,y,z,t)$, and  the  time-dependent streamwise mean  flow, $\mathbf{U}(y,z,t)$.  Equation \eqref{eqn:RNL-perturb} can be linearized around $\mathbf{U}(y,z,t)$ to yield
\begin{equation}
\mathbf{u}_t =  A(\mathbf{U})  \mathbf{u}+\mathbf{e},
\label{eqn:Au}
 \end{equation}
where  $A(\mathbf{U})$ is the associated linear operator. \iffalse \tc{red}{\st{Here, ${A} (\mathbf{U})\mathbf{u}$ is a bilinear function of the time-varying streamwise mean flow, $\mathbf{U}(t)$, and the perturbation velocity, $\mathbf{u}$.}} \tc{blue}{The expression \eqref{eqn:Au} is derived from \eqref{eqn:RNL-perturb} by writing the perturbation in terms of the normal velocity $v$ and normal vorticity $\eta = \partial_z u - \partial_x w$, see e.g.~\cite{Benney-Gustavsson-1981,Schmid-Henningson-2001} eliminating the pressure and then transforming the system back into the $(u,v,w)$ variables as in \citet{Jovanovic-Bamieh-2005}.} \fi

The closely related S3T system is derived by making the ergodic assumption of equating the streamwise average  with the ensemble average over  realizations of the stochastic excitation $\mathbf{\epsilon}$ in \eqref{eq:NSp}.  The S3T system is a second order closure of the NS equations in \eqref{eq:NSE0}, in which the first order cumulant is  $\mathbf{U}$ and the second order cumulant is  the spatial covariance $C$ between any two points $\mathbf{x}_1$ and $\mathbf{x}_2$. We refer to the resulting closed system of equations as the statistical state dynamics of the flow:
\begin{subequations}
 \label{eqn:S3T}
 \begin{eqnarray}
{\mathbf{U}_t} = \mathbf{U} \cdot \nabla \mathbf{U}
 + \nabla \mathbf{P} -  \frac{1}{R}\Delta \mathbf{U}  + \mathcal{L}C \label{eqn:mean_flow-S3T} \\
 {C_t} = \left ( A _1(\mathbf{U})+   A _2(\mathbf{U} ) \right )   C + Q \label{eqn:perturb-S3T}.
\end{eqnarray}
\end{subequations}
Here,  $Q$ is the second order covariance of the stochastic excitation, which is assumed to be temporally delta correlated. $\mathcal{L}C$ denotes the divergence of the streamwise
Reynolds stresses expressed as a linear function of the covariance $C$. The expression $\mathbf{A}_1(\mathbf{U}) C$ accounts for the contribution to the time rate of change of the covariance arising from the action of the operator $A(\mathbf{U})$ evaluated at point $\mathbf{x}_1$  on the corresponding  component of $C$. A similar relation holds for
 $\mathbf{A}_2(\mathbf{U}) C$. Further details regarding equations \eqref{eqn:Au} and  \eqref{eqn:S3T} are provided in Appendix A.

 In isolation, the mean flow dynamics  \eqref{eqn:mean_flow-S3T} define a streamwise constant or 2D/3C model of the flow field~\cite{BobbaThesis,Gayme-etal-2010} forced  by the Reynolds stress divergence specified by $\mathcal{L}C$. The S3T system \eqref{eqn:S3T} describes the statistical state dynamics closed at second order, which has been shown to be sufficiently comprehensive to allow identification of statistical equilibria of turbulent flows and  permit analysis of their stability~\cite{Farrell-Ioannou-2012}. S3T provides an attractive theoretical framework for studying turbulence through analysis of its underlying statistical mean state dynamics. However, it has the perturbation covariance as a variable and as a result it becomes computationally intractable for high dimensional systems.

 The RNL model shares the dynamical restrictions of S3T, and therefore its properties can be directly related to  the S3T system. Since the RNL model in \eqref{eqn:RNL} uses a single realization of the infinite ensemble that makes up S3T dynamics to approximate the ensemble covariance,  it avoids explicit time integration of the perturbation covariance equation and facilitates computationally efficient studies of the S3T system dynamics. The RNL system also has the advantage that it can be easily implemented by restricting a DNS code to the RNL dynamics in \eqref{eqn:RNL}.\iffalse The perturbation covariance has dimension O($N^2$) for a system of dimension O($N$) and is only directly integrable for low order systems.\fi 

In this paper we consider the unforced RNL system, which corresponds to setting $\mathbf{e}=0$ in \eqref{eqn:RNL-perturb}. This system models the dynamics occurring after an initial transient phase  during which an excitation has been applied  to initiate turbulence. We demonstrate that subsequent to this transient phase the RNL system supports turbulence that closely resembles that seen in DNS of fully developed turbulence in plane Couette flow.

\begin{table*}
\centering
\caption{\label{table:geometry}Geometry for numerical simulations. $x$, $y$, and $z$ define the computational domain. $N_x$, $N_y$ and $N_z$ are the number of grid points in their respective dimensions. $M_x$, and $M_z$ are the number of Fourier modes used after dealiasing and $M_y$ is the number of Chebyshev modes used in each simulation.}
\begin{tabular}{p{1.5cm}p{1.25cm}p{1.25cm}p{1.5cm}p{3cm}p{2.5cm}} \hline
& \parbox[c]{1cm}{$x$} &\parbox[c]{1cm}{$y$} &\parbox[c]{1cm}{$z$} & $N_x \times N_y \times N_z$ & $M_x \times M_y \times M_z$  \\ \hline
DNS  & $[0,4\pi]$ & $[-1,1]$ & $[0,4\pi]$ & $128 \times 65 \times 128$  & $83\times 65 \times 41$ \\
RNL  & $[0,4\pi]$ & $[-1,1]$ & $[0,4\pi]$ & \hspace{0.7 mm} $16 \times 65 \times 128$ & \hspace{0.7 mm} $9 \times 65 \times 41$ \\
2D/3C  &  & $[-1,1]$ & $[0,4\pi]$ & \hspace{10 mm} $65 \times 128$ & \hspace{8 mm} $65 \times 41$ \\ \hline
\end{tabular}
\end{table*}

%========================================================================
\subsection{Numerical method}
\label{sec:numerics}

The numerical simulations in this paper were carried out using a spectral code based on the Channelflow NS equations solver \cite{Gibson-2007}. The time integration uses a third order multistep semi-implicit Adams-Bashforth/backward-differentiation scheme that is detailed in \cite{Peyret2002}. The discretization time step is automatically adjusted such that the Courant-Friedrichs-Lewy (CFL) number is kept between $0.05$ and $0.2$. The spatial derivatives employ Chebyshev polynomials in the wall-normal ($y$) direction and Fourier series expansions in streamwise ($x$) and spanwise ($z$) directions \citep{Canuto1988}. No-slip boundary conditions are employed at the walls for the $y$ component and periodic boundary conditions are used in the $x$ and $z$ directions for all of the velocity fields. Aliasing errors from the Fourier transforms are removed using the 3/2-rule, as detailed in \cite{Zang1985}. A zero pressure gradient was imposed in all simulations. Table \ref{table:geometry} provides the dimensions of the computational box, the number of grid points, and the number of spectral modes for the DNS and simulations of the RNL and 2D/3C systems. In the DNS we use the stochastic excitation $\varepsilon$ in \eqref{eq:NSE0} only to initiate turbulence. In order to perform the RNL computations the DNS code was restricted to the dynamics of \eqref{eqn:RNL} with $\mathbf{e}=0$, subsequent to the establishment of the turbulent state. For simulations of the 2D/3C system, the time varying streamwise mean flow in the perturbation dynamics was eliminated by replacing the term $\mathbf{U} \cdot \nabla \mathbf{u} + \mathbf{u} \cdot \nabla \mathbf{ U}$ on the right-hand-side of \eqref{eqn:RNL-perturb} with $\mathbf{U_{lam}} \cdot \nabla \mathbf{u} + \mathbf{u} \cdot \nabla \mathbf{ U_{lam}}$, where $\mathbf{U_{lam}}= (U(y), 0, 0)$ defines the laminar velocity profile for plane Couette flow  with $U(y)=y$.
\section{Results}
\label{sec:results}

In this section we compare simulations of the RNL system \eqref{eqn:RNL} to DNS of fully developed turbulence in
plane Couette flow. The geometry and resolution for each of the DNS and RNL cases in this section are given in Table \ref{table:geometry}. Turbulence is initiated by applying the stochastic excitation $\mathbf{\epsilon}$ in \eqref{eq:NSp} for the DNS cases and $\mathbf{e}$ in \eqref{eqn:RNL-perturb} for the RNL simulations over the interval $t \in [0,500]$, where $t$ represents convective time units. All of the results reported in this section are for $R=1000$.

\begin{figure}
\subfloat[\label{fig:TurbulentProfile}]{
\includegraphics[width = 0.45\textwidth,clip=]{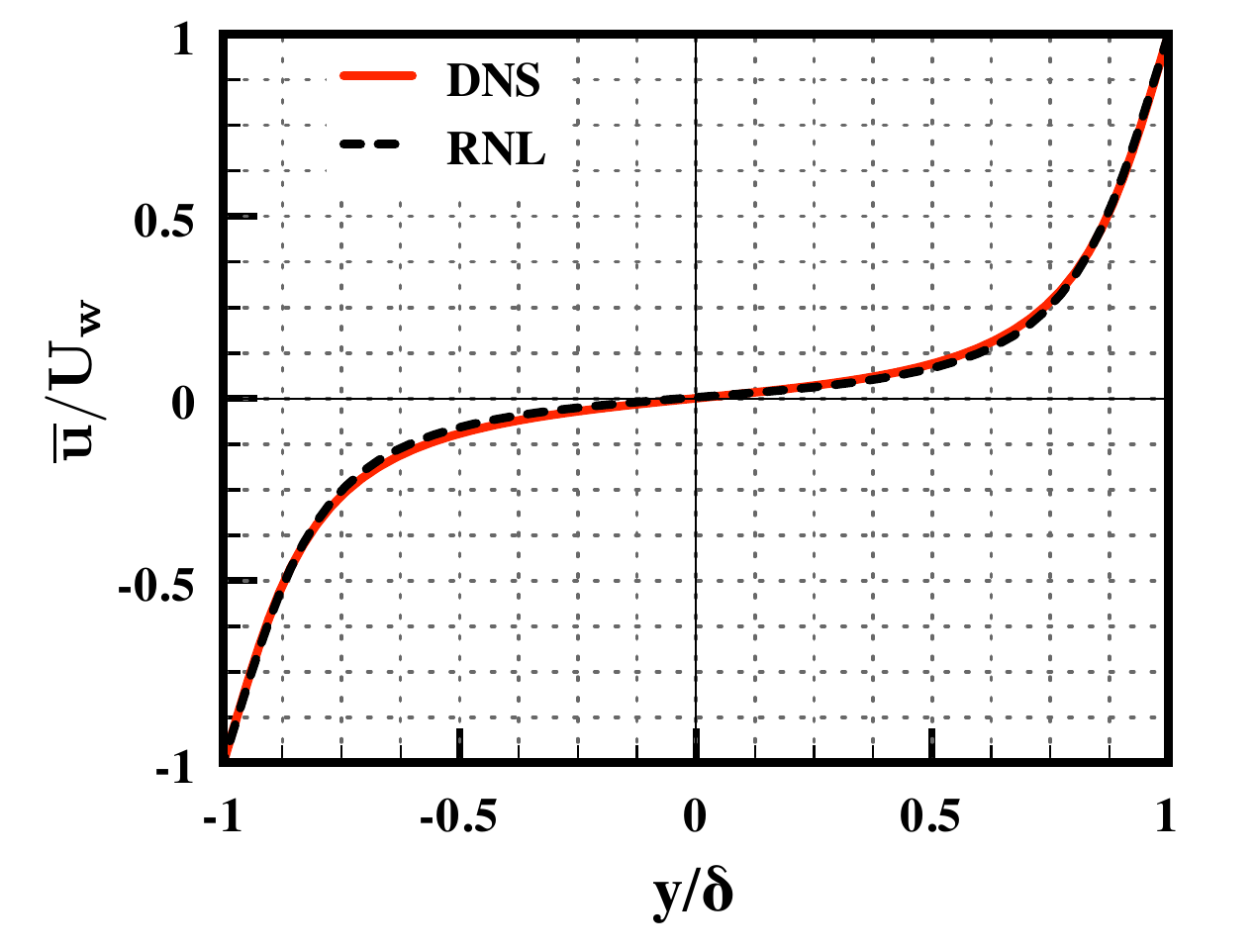}}
\subfloat[\label{fig:WallUnitTurbulentProfile}]{
\includegraphics[width = 0.45\textwidth,clip=]{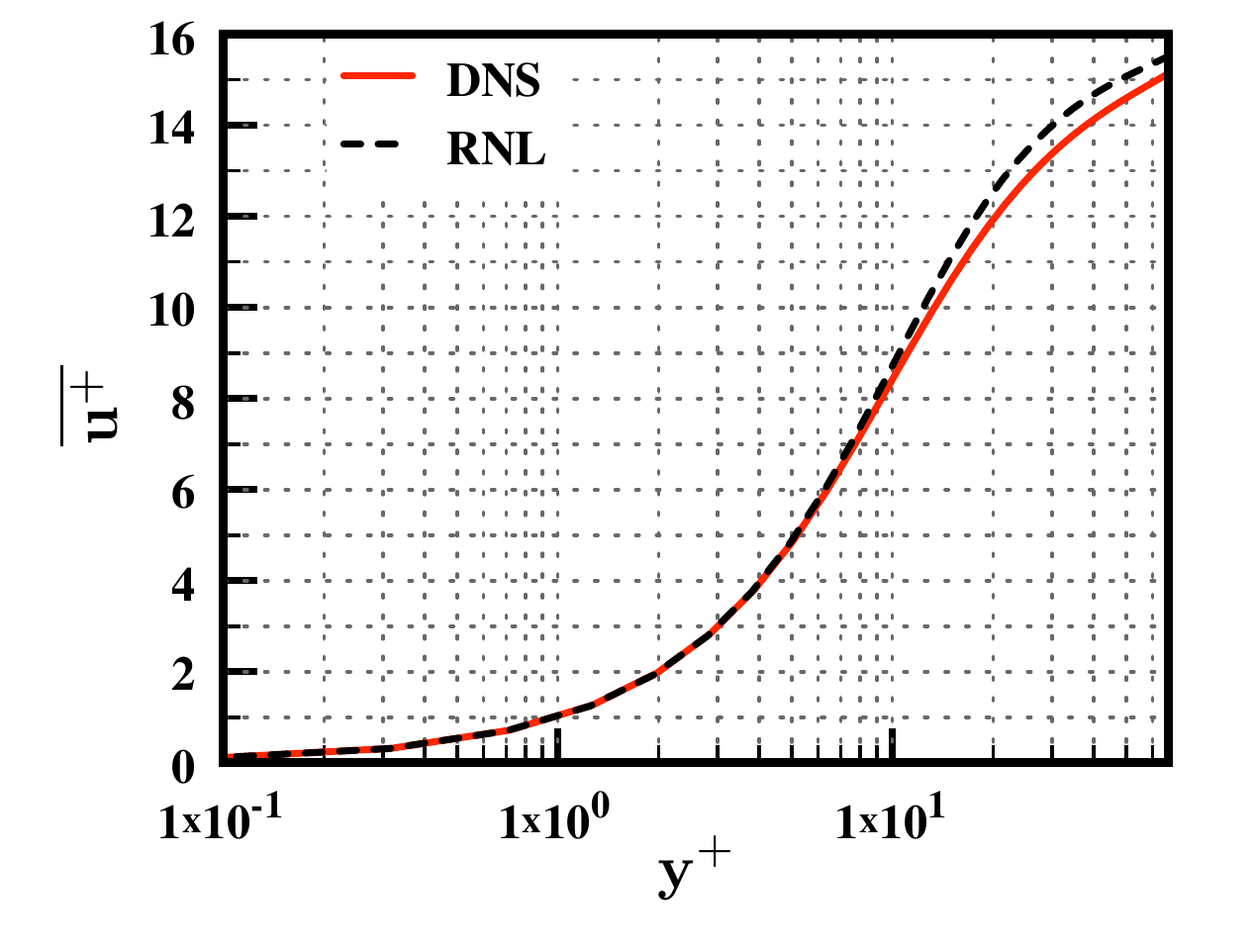} }
\caption{Turbulent mean velocity profiles (based on streamwise, spanwise and time averages) in (a) geometric units and (b) wall units obtained from DNS (red solid line) and unforced RNL simulations (black dashed line).}
  \label{fig:turbulentprofiles}
 \end{figure}

The turbulent mean velocity profile obtained from the DNS is compared to that obtained using the unforced RNL system in Figure \ref{fig:TurbulentProfile}. Figure \ref{fig:WallUnitTurbulentProfile} provides a comparison of the same data in wall units, $u^+$ = $u/u_{\tau}$ and $y^+$ = $y \hspace*{0.3 mm} u_{\tau}/\nu$ with friction velocity  $u_{\tau}$ = $\sqrt{\tau_w/\rho}$, $Re_{\tau}$ = $u_{\tau}\delta/\nu$ and $\nu$ = $1/R$. These wall unit values for the DNS results in Figure \ref{fig:WallUnitTurbulentProfile} are ${Re_{\tau}}$ = $66.2$ and ${u_{\tau}/U_w}$ =  $6.62 \times 10^{-2}$ and those corresponding to the RNL simulation are ${Re_{\tau}}$ = $64.9$ and ${u_{\tau}/U_w}$ =  $6.49 \times 10^{-2}$.  Figure \ref{fig:turbulentprofiles} illustrates good agreement between the turbulent mean velocity profile obtained from the RNL simulation and that obtained from the DNS, which is consistent with recent studies~\cite{Farrell-etal-2012-ctr}.

\begin{figure*}[!h]
  \centering
  \subfloat[{\bf \ DNS} \label{fig:dns_contour}]{
   \includegraphics[width = 1\textwidth,clip]{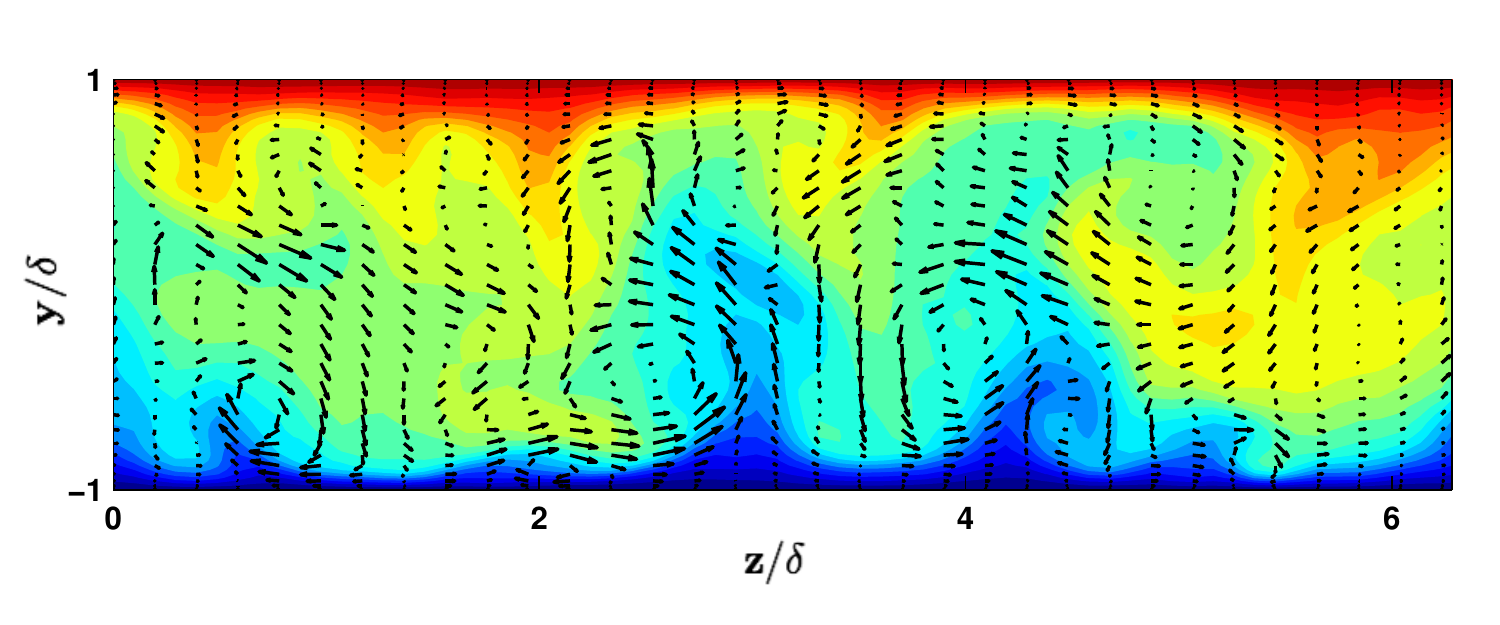}
   } \\
  \subfloat[{\bf \ RNL} \label{fig:RNL_contour}]{
  \includegraphics[width = 1\textwidth,clip]{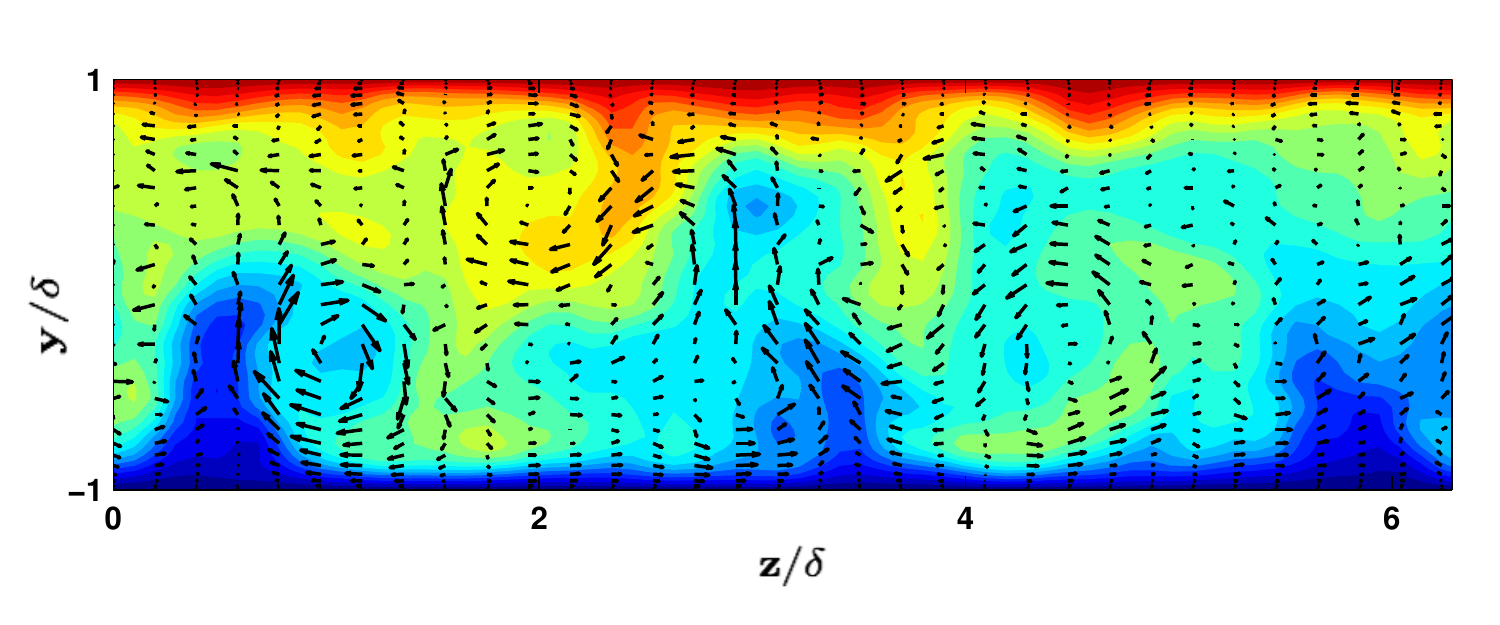}
   }
   \caption{A $y$-$z$ plane cross-section of the streamwise constant structure (component $k_x$ = 0) of the flow at a single snapshot in time
   for a (a) DNS and a (b) RNL simulation. Both panels show contours of the
   streamwise component of the mean flow $U$ with the velocity vectors of ($V$,$W$) superimposed. The RNL is self-sustaining ($\mathbf{e}$ = 0) for the time shown}
\label{fig:contours}
\end{figure*}

\begin{figure*}[!h]
  \centering
  \subfloat[{\bf \ DNS}\label{fig:dns_perspective}]{
   \includegraphics[width = .9\textwidth,clip]{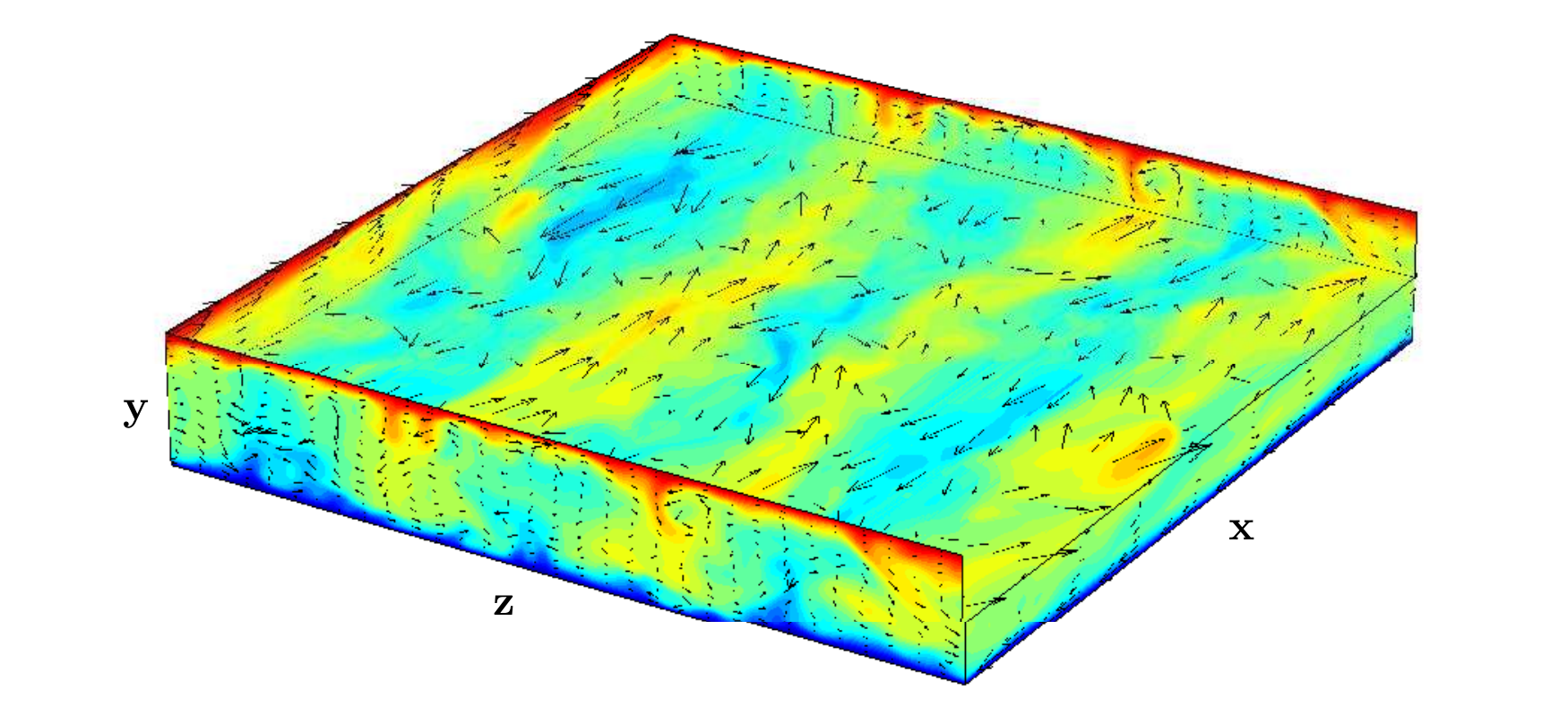}
    } \\
  \subfloat[][{\bf \ RNL}\label{fig:RNL_perspective}]{
  \includegraphics[width = .9\textwidth,clip]{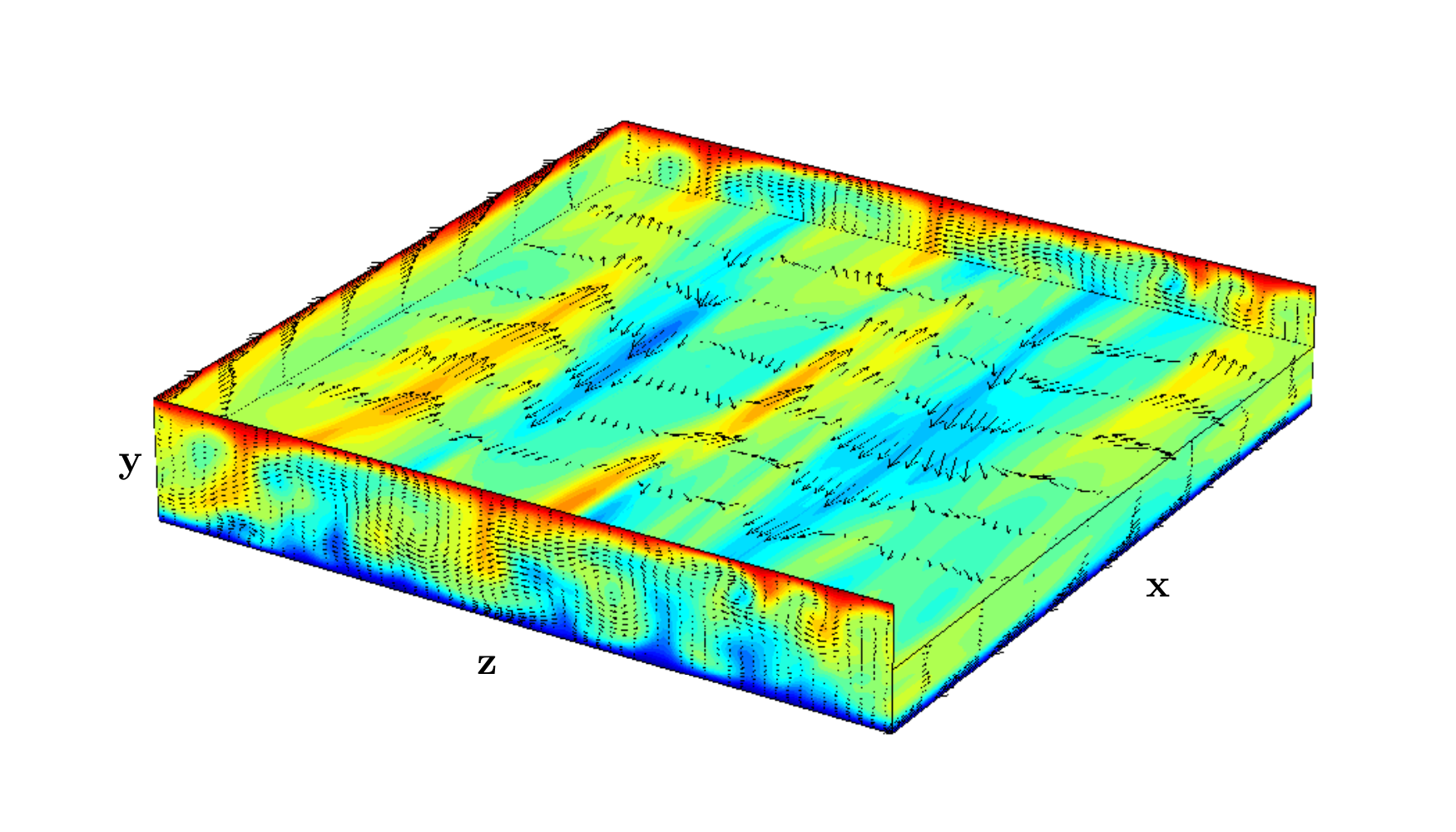}
    }
   \caption{3D perspective plots of the flow at a single snapshot in time for (a) a DNS, and (b) a RNL simulation  Both images show contours of the streamwise component of the mean flow, $U$. The superimposed vectors represent the in-plane velocity vectors for the respective panel (i.e.:  $(V,W)$ for the $y-z$ panels and $(U,W)$ for $x-z$ panels). The central $x-z$ panel shows the flow at the $y = 0$ mid-plane of the system. The geometry of the systems are detailed in Table \ref{table:geometry}.}
\label{fig:perspective}
\end{figure*}

Instantaneous snapshots of the turbulent velocity fields from the DNS and the RNL simulation are displayed in Figures \ref{fig:contours} and \ref{fig:perspective}. Figure \ref{fig:contours} shows contour plots of the $U$ velocity field with the $V$, $W$ vector fields superimposed. The large-scale roll structures characteristic of turbulent flow  are evident in both simulations.  Figure \ref{fig:perspective}  illustrates the three-dimensional structure of the streamwise component of the velocity field. Figures \ref{fig:contours} and \ref{fig:perspective} demonstrate that the DNS and the RNL system produce similar structures including comparable streak spacing and associated roll circulations. The visual similarity of the roll and streak structures produced by the DNS and RNL simulations implies that the RNL accurately captures these fundamental features of fully developed turbulent plane Couette flow.


\begin{figure}
\subfloat[]{\label{fig:uuWallUnits}
\includegraphics[width = 0.45\textwidth,clip=]{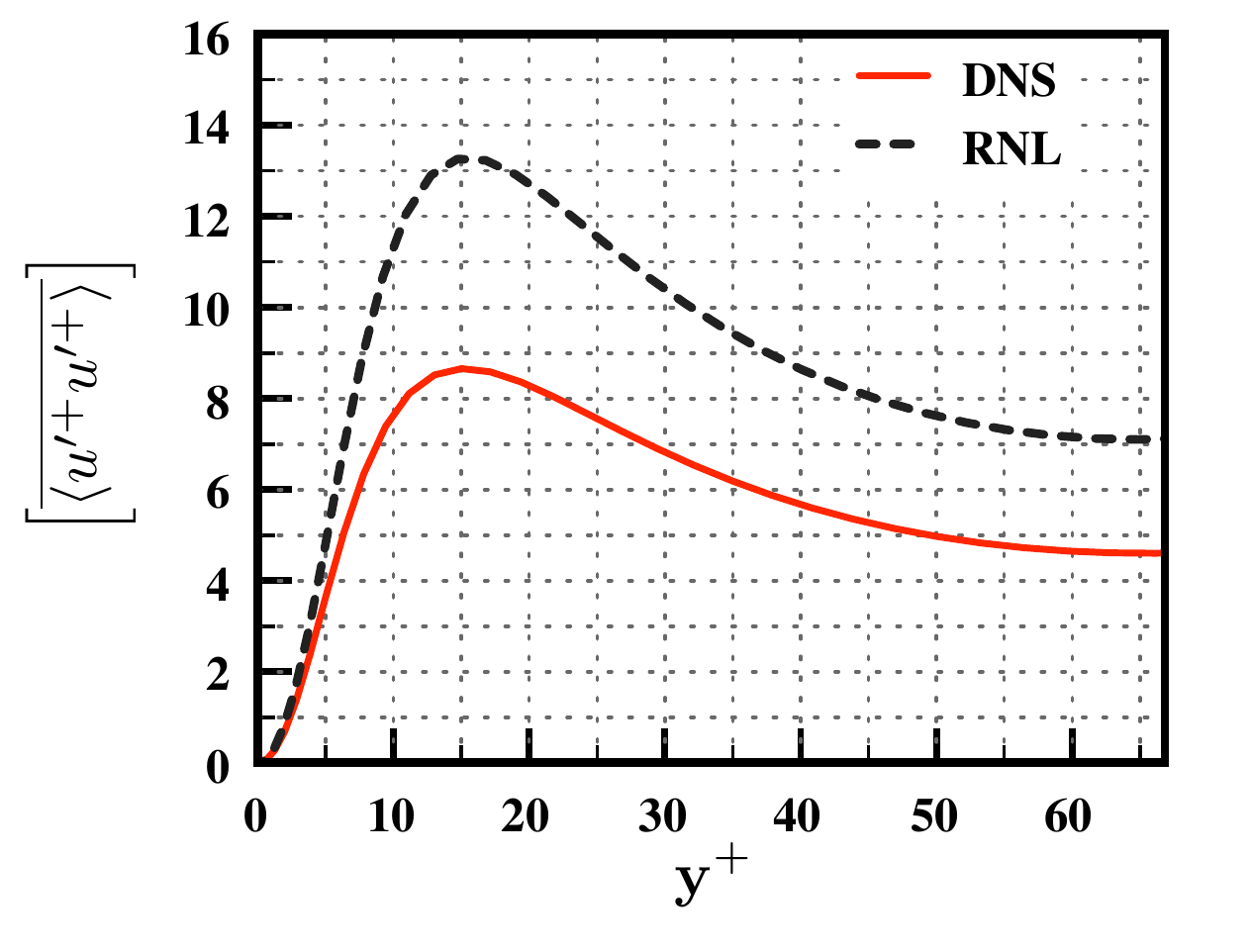}}
\subfloat[]{\label{fig:uvWallUnits}
\includegraphics[width = 0.45\textwidth,clip=]{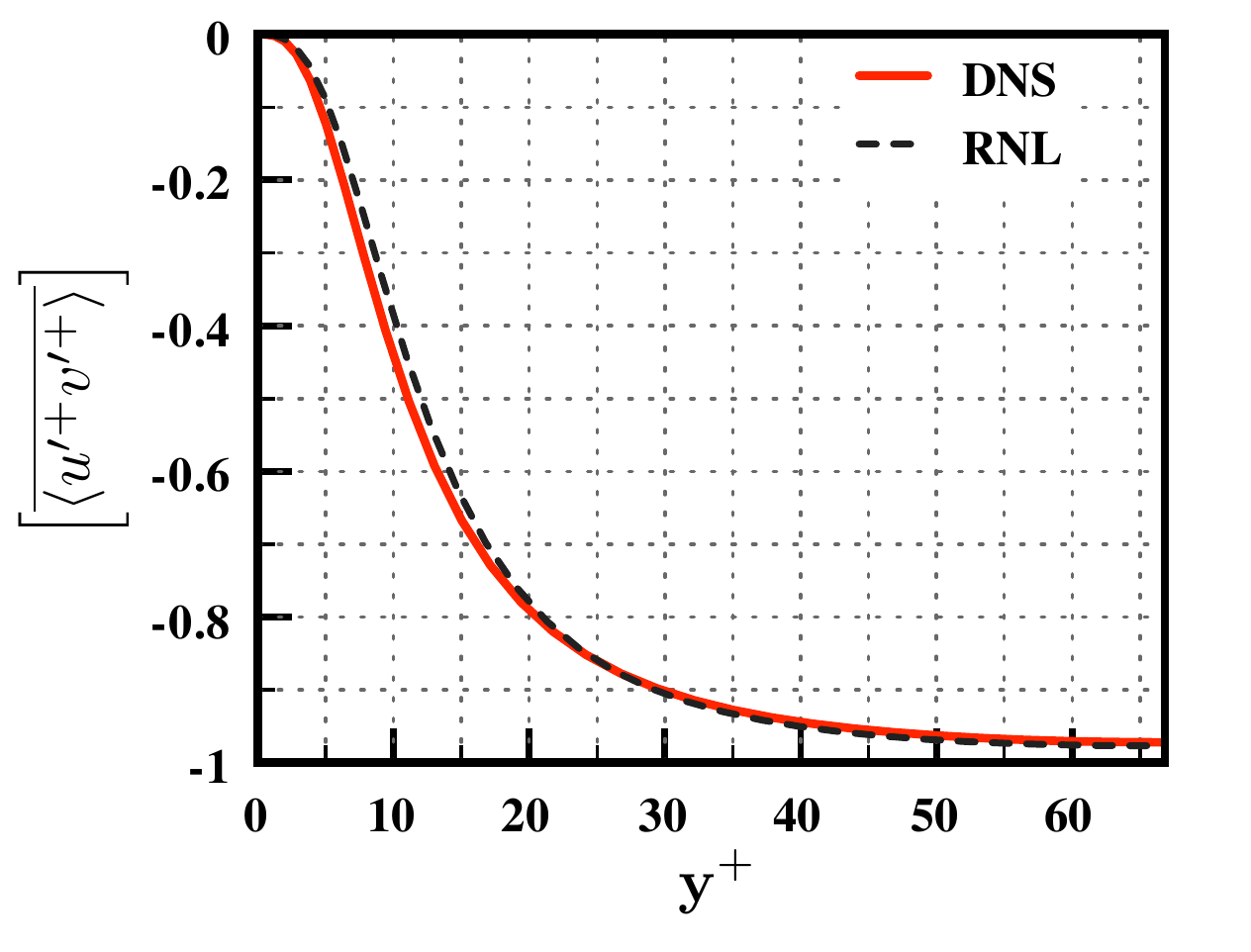} }
\caption{ Reynolds stresses (a) $\left[\overline{\langle u'^+u'^+\rangle }\right]$ and (b) $\left[\overline{\langle u'^+v'^+ \rangle} \right]$ obtained from a DNS and a RNL simulation. There is no stochastic excitation  applied to the DNS or the RNL simulation during the time interval used to generate these profiles, (i.e. the RNL simulation is in the self-sustaining state)  }
  \label{fig:turbstats}
\end{figure}

Figure~\ref{fig:turbstats} shows the corresponding time averaged Reynolds stresses, $\left[\overline{\langle u'^+u'^+\rangle }\right]$ and $\left[\overline{\langle u'^+v'^+ \rangle} \right]$, where the streamwise fluctuations, $u'$, are defined as $u'$ =  $u_T - \overline{u_T }$ and the wall-normal fluctuations , $v'$, are defined as $v'$ =  $v_T - \overline{v_T}$.  $u'^+$ and $v'^+$ designate these fluctuations scaled by $u_{\tau}$, such that $u'^+$ =  $u'/u_{\tau}$ and $v'^+$ = $v'/u_{\tau}$. These figures illustrate close agreement between the $\overline{u'v'}$ Reynolds stress obtained from the RNL and DNS. However, the $\overline{u'u'}$ component has a higher magnitude in the RNL than in the DNS. As shown in Figure \ref{fig:WallUnitTurbulentProfile}, the turbulent flow supported by DNS and the RNL simulation exhibit nearly identical shear at the boundary. Therefore, the average energy input and by consistency the dissipation must be the same in these simulations.  However, the RNL maintains a smaller number of streamwise Fourier components than the DNS~\cite{Thomas-etal-preprint-2}. The result in Figure \ref{fig:uuWallUnits} is thus consistent with the smaller number of streamwise components supported by the RNL system producing the same dissipation as the DNS. This aspect of the dynamics is a subject of continuing investigation~\cite{Thomas-etal-preprint-2}.

\begin{figure}
\subfloat[]{\label{fig:departure}
\includegraphics[width = 0.45\textwidth,clip=]{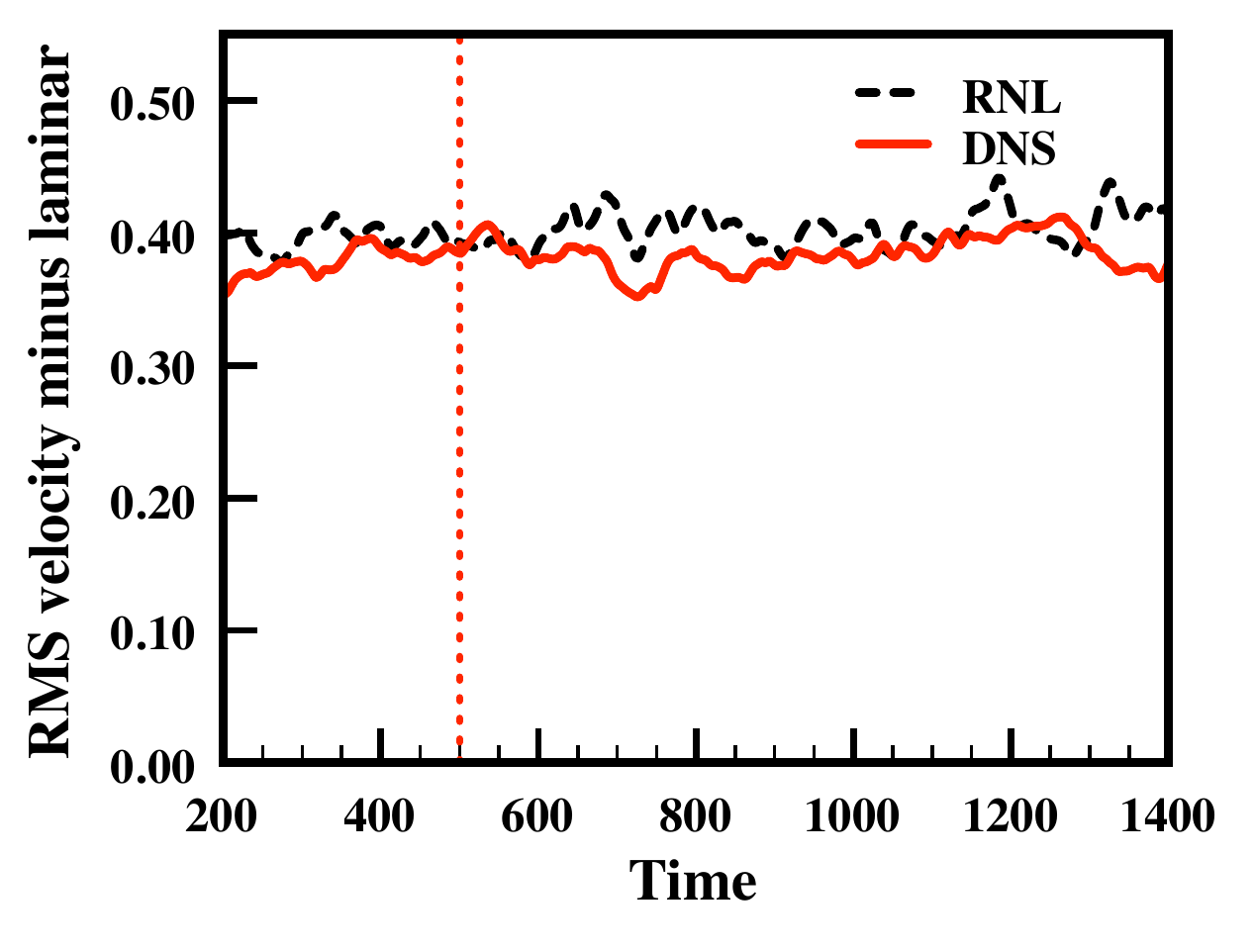}}
\subfloat[]{\label{fig:perturbation}
\includegraphics[width = 0.45\textwidth,clip=]{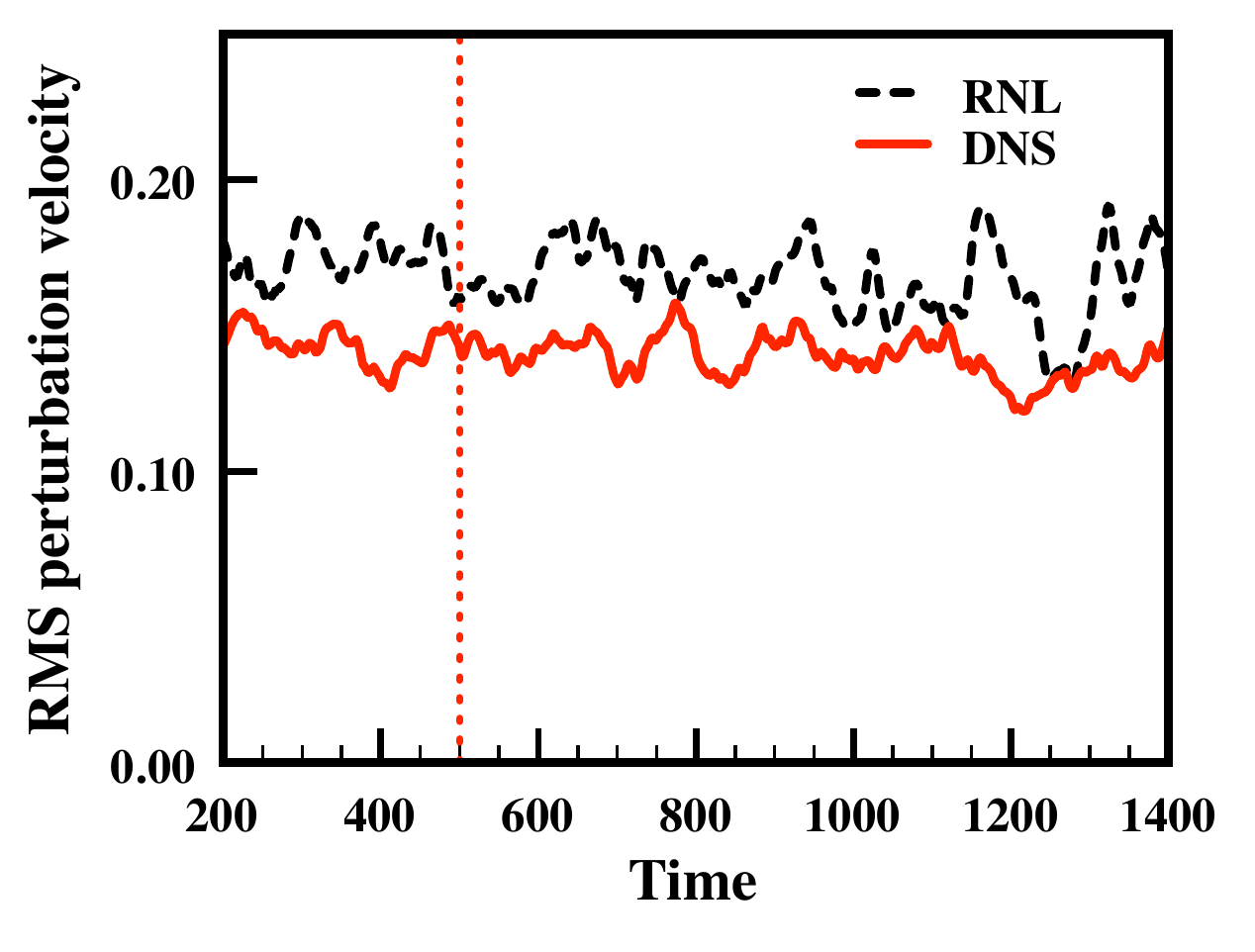} } \\
\subfloat[]{\label{fig:Re_tau}
\includegraphics[width = 0.45\textwidth,clip=]{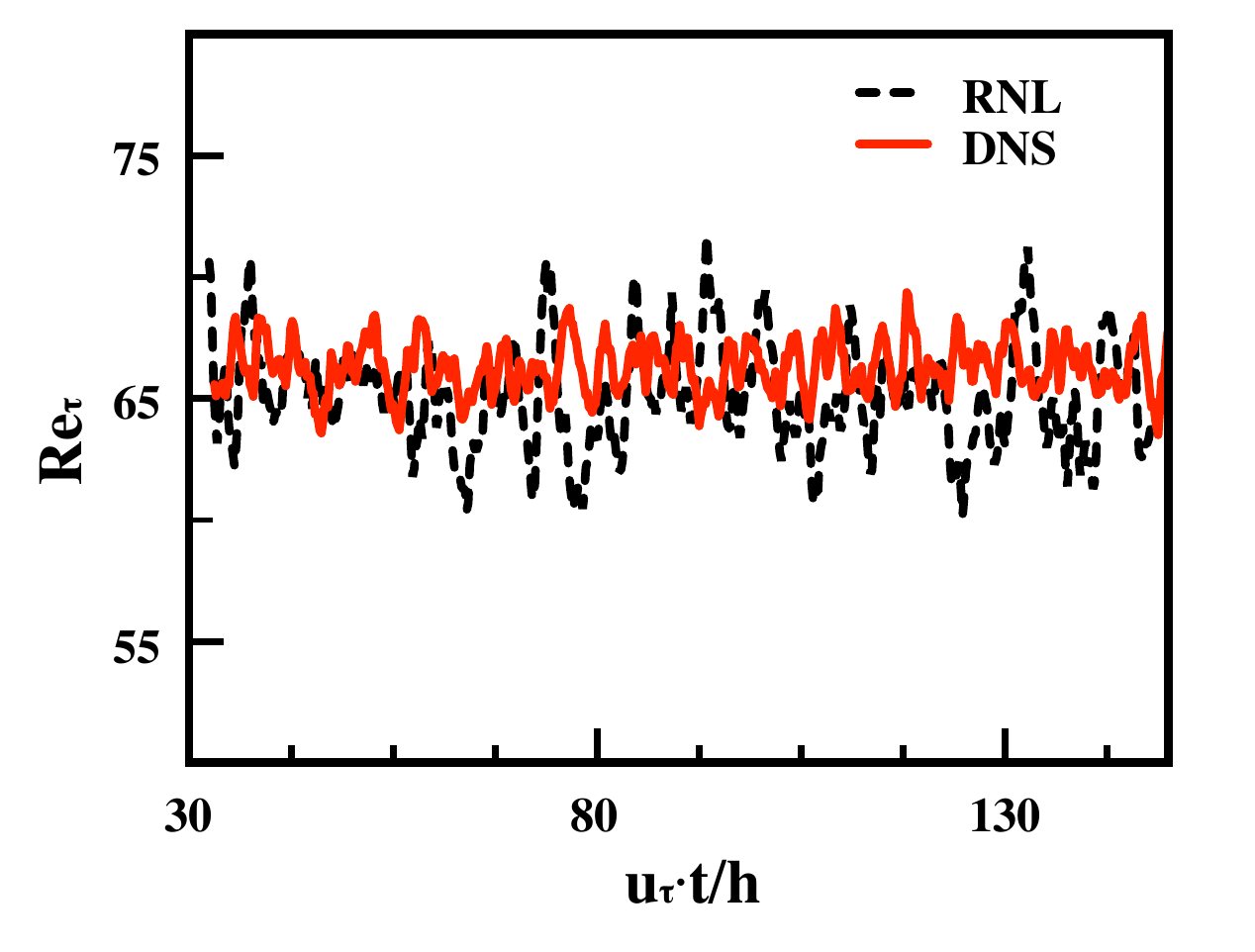} }
            \caption{ (a) RMS velocity minus laminar, $\sqrt{\left(\left(\mathbf{U}+\mathbf{u}\right) - \mathbf{U}_{\mathbf{lam}}\right)^2}$, (b) RMS perturbation velocity $\sqrt{u^2+v^2+w^2}$. The external excitation was stopped at $t=500$. The behavior of the RNL simulation for $t>500$ is similar to that of the DNS, indicating that the RNL  simulation undergoes the SSP described in \cite{Farrell-Ioannou-2012},  and (c) $Re_\tau$ versus dimensionless time, $u_{\tau}t/ h$ obtained from DNS (red solid line) and the RNL (black dashed line) simulation. For DNS, $Re_{\tau}$ = $66.2$ and $u_{\tau}/U_w$ =  $6.62 \times 10^{-2}$. In the RNL simulation, ${Re_{\tau}}$ = $64.9$ and ${u_{\tau}/U_w}$ =  $6.49 \times 10^{-2}$}
  \label{ffig:velocity profiles}
\end{figure}

Figures \ref{fig:departure} and \ref{fig:perturbation} shows close agreement in the root-mean-square (RMS) velocity departure from laminar, defined as $\sqrt{\left(\mathbf{U}+\mathbf{u} - \mathbf{U}_{\mathrm{lam}}\right)^2}$, and the RMS perturbation velocity, $\sqrt{u^2 + v^2 + w^2}$, between the RNL simulation and DNS. In particular, the RNL simulation maintains the same behavior after the initial forcing is removed (for $t > 500$). In Figure \ref{fig:Re_tau} the friction Reynolds number, $Re_{\tau}$, is displayed as a function of dimensionless time, $u_{\tau}t/h$. The time interval in Figure \ref{fig:Re_tau} corresponds to $t \in [1000,6000]$, which verifies that the RNL system maintains turbulence for an extended interval of time. The RNL simulation thus exhibits both self-sustaining behavior and dissipation comparable to that in DNS.

The results shown in Figure \ref{ffig:velocity profiles} indicate  that the self-sustaining behavior captured by the RNL system supports a SSP similar to that previously seen in the related S3T system~\cite{Farrell-Ioannou-2012}. Previous analysis of the S3T system established that this SSP is due to the coupling between the streamwise mean flow and the perturbations. In particular, the roll circulations are driven by the Reynolds stresses arising from the perturbations.~\cite{Farrell-Ioannou-2012}.  In turn, maintenance of this perturbation field has been shown to result from a parametric non-normal growth process arising from the interaction between the time-dependent streak (resulting from the roll circulations) and the perturbation field~\cite{Farrell-Ioannou-1996b,Farrell-Ioannou-1999, Farrell-Ioannou-2012}.  The similarity between the SSP operating in the S3T/RNL system with that in DNS provides evidence that the same mechanism underlies the SSP operating in these systems.

\section{Comparison of RNL and 2D/3C Models}
\label{sec:2d3c}
We now verify the fundamental role of the coupling between  the mean flow  equation \eqref{eqn:RNL-mean}  and  the perturbation equation \eqref{eqn:RNL-perturb} in the maintenance of turbulence in the RNL system by comparing the RNL and 2D/3C models \cite{Gayme-etal-2010}. The 2D/3C system can be obtained from \eqref{eqn:S3T} by replacing the time varying mean flow $\mathbf{U}$ in \eqref{eqn:perturb-S3T} with the time-invariant laminar Couette flow $\mathbf{U}_{\mathbf{lam}}=U(y)$. The interaction whereby the mean flow influences the perturbations has thus been eliminated and the perturbation
covariance evolves under stochastic forcing of the stable  $A(\mathbf{U}_{\mathbf{lam}})$ for all times. In this case the mean flow dynamics represent a forced streamwise constant  (2D/3C) system given by
\begin{subequations}
 \label{eqn:2D3C}
 \begin{eqnarray}
 \mathbf{U} \cdot \nabla \mathbf{U}
 + \nabla \mathbf{P} -  \frac{1}{R}\Delta \mathbf{U}  =- \mathcal{L}C^{\infty}\label{eqn:mean_flow2D3C} \\
  \left ( A_1(\mathbf{U}_{lam}) + A_2 (\mathbf{U}_{lam} ) \right ) C^{\infty} = -  Q ~, \label{eqn:perturb2D3C}
\end{eqnarray}
\end{subequations}
where $C^{\infty}$ denotes the asymptotic equilibrium covariance.

\begin{figure}
\subfloat[]{\label{fig:TurbulentProfile with 2D3C}
\includegraphics[width = 0.45\textwidth,clip=]{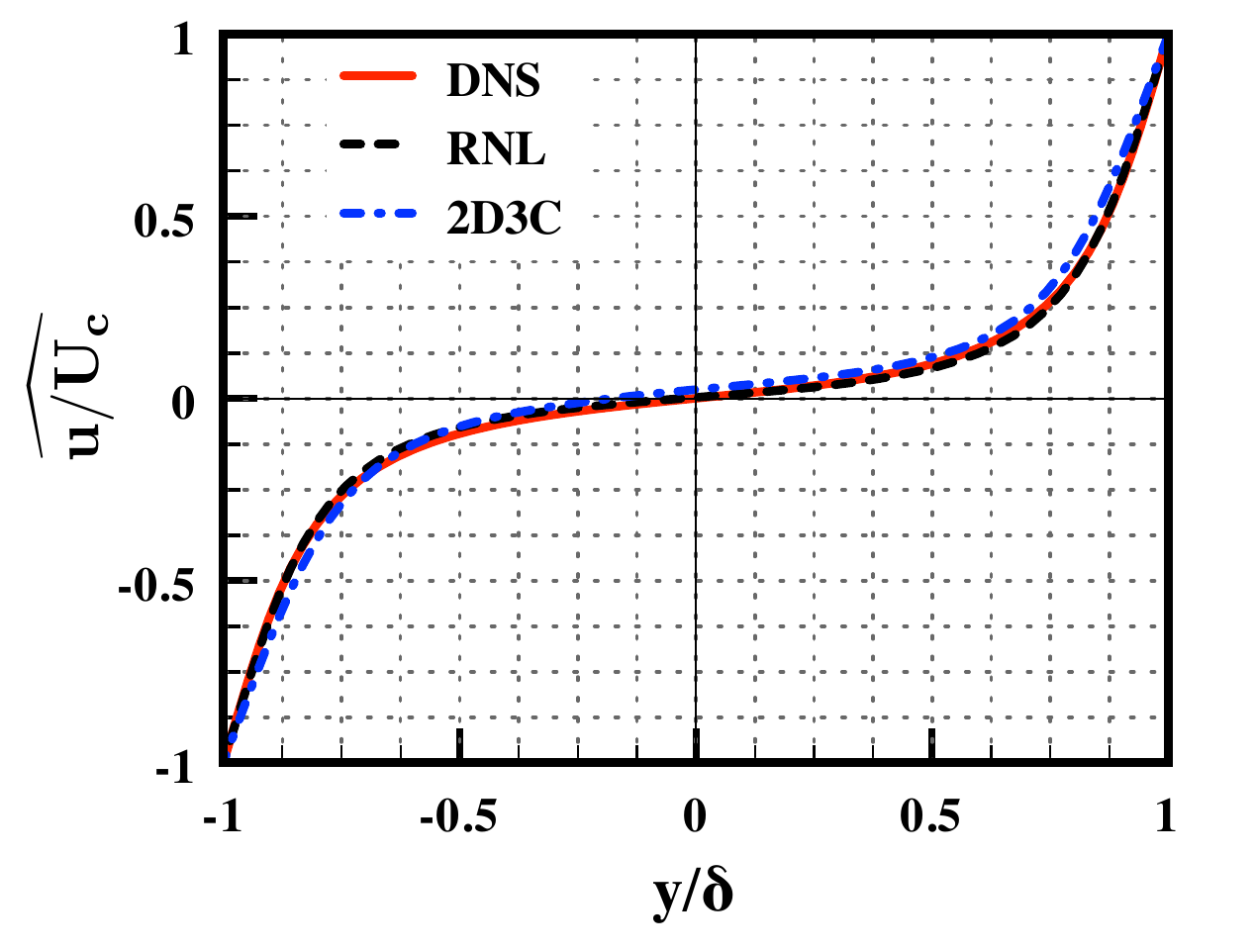}}
\subfloat[]{\label{fig:streak with 2D3C}
\includegraphics[width = 0.45\textwidth,clip=]{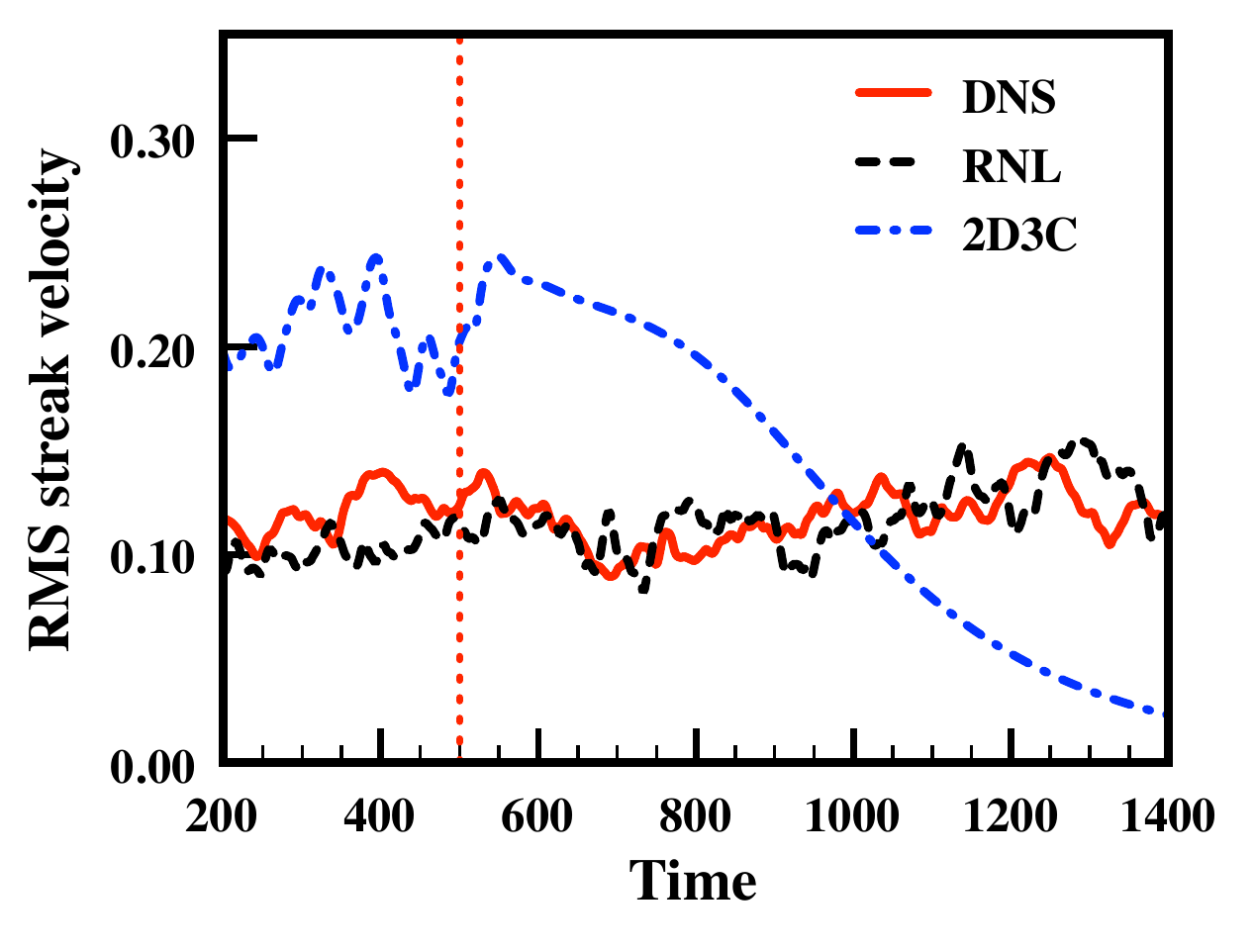} }
\caption{(a)  Turbulent mean velocity profiles (based on a streamwise, spanwise and time averages)  obtained from  DNS (red solid line),  and simulations of the RNL (black dashed line) and 2D/3C systems (blue dashed-dot line).  There is no stochastic excitation applied to the DNS or the RNL simulation during the time interval used to generate the profile, whereas the 2D/3C simulation was continuously forced at $\mathbf{e}$ = 0.030. (b) The RMS streak velocity $\sqrt{U_s^2}$  obtained from DNS (red solid line) as well as the RNL (black dashed line) and 2D/3C simulations (blue dashed-dot line) where stochastic excitation was applied for $t \in[0,500]$, i.e., the excitation was stopped at $t=500$, which is indicated by the vertical red dotted line.  }
  \label{fig:2D3C}
\end{figure}

Figure \ref{fig:TurbulentProfile with 2D3C} shows the same mean velocity profiles as in Figure \ref{fig:TurbulentProfile} along with that obtained using a stochastically forced 2D/3C model \cite{Gayme-etal-2010}. This plot demonstrates close correspondence between the mean velocity profiles obtained from DNS and simulations of the 2D/3C and RNL systems. Figure \ref{fig:streak with 2D3C} shows the time evolution of the RMS streak velocity from simulations of the 2D/3C and RNL systems and DNS. Figure \ref{fig:streak with 2D3C} shows that the streak in the 2D/3C model gradually decays to zero after the external excitation is removed at $t=500$. This figure demonstrates that a stochastically forced 2D/3C model captures the turbulent mean flow profile, but cannot maintain turbulence without stochastic excitation, see e.g. \cite{BobbaThesis}. \iffalse The 2D/3C system also returns to the laminar flow after the excitation. \fi

\begin{figure}
       \includegraphics[width = 0.4\textwidth,clip=]{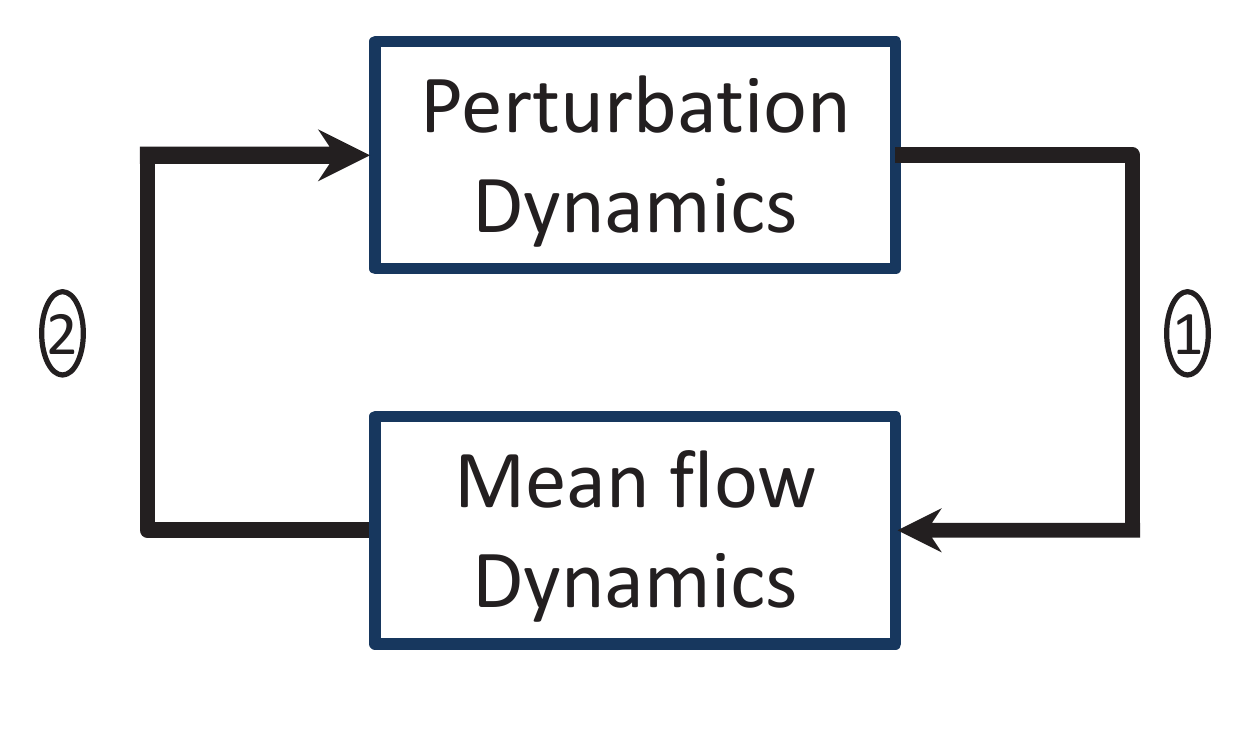}
      \caption{\label{fig:block_diagram} In both the 2D/3C and the unforced RNL model \eqref{eqn:RNL} (and its associated ensemble mean  S3T model) the perturbations ($\mathbf{u}$) influence the dynamics of the mean flow ($\mathbf{U}$). This coupling is denoted pathway \textcircled{1} in the block diagram. The S3T model augments the 2D/3C formulation with feedback from the mean flow to the perturbation dynamics, which is illustrated through pathway \textcircled{2}. }
        \end{figure}

The critical difference between the 2D/3C and RNL systems is that the 2D/3C model lacks the interaction of the time-varying mean flow \eqref{eqn:RNL-mean} with the perturbation dynamics \eqref{eqn:RNL-perturb}. This difference is summarized by the block diagram  in Figure \ref{fig:block_diagram}. Both of these models include pathway \textcircled{1} in which the perturbations ($\mathbf{u}$) influence the dynamics of the mean flow ($\mathbf{U}$). However, the RNL system (and its associated ensemble mean S3T model) also includes the feedback pathway \textcircled{2}, from the mean flow to the perturbation dynamics. In Figure \ref{fig:streak with 2D3C} the effect of this feedback from the mean flow to the perturbations, pathway \textcircled{2} in Figure \ref{fig:block_diagram}, is seen to be critical for capturing the mechanism of the SSP maintaining the turbulent state. As shown in the previous section and in Figure \ref{fig:2D3C}, turbulence in the RNL system self-sustains (i.e.,  is maintained in the absence of stochastic excitation). Turbulence maintained by the same mechanism was seen previously in the S3T system in a  minimal channel study~\cite{Farrell-Ioannou-2012}.

In addition to being necessary to produce self-sustaining turbulence, the feedback from the mean flow to the perturbations produces streaks that are quantitatively and qualitatively similar to those observed in DNS, and notably more accurate than those obtained in the 2D/3C system. This result is consistent with the fact that in the 2D/3C model the streak is not regulated by feedback from the mean flow to the perturbation field, pathway \textcircled{2} in Figure \ref{fig:block_diagram}. Therefore, understanding how the roll and streak structure and the mechanism by which it is maintained in a statistical steady state requires a model that includes feedback from the streamwise constant mean flow to the streamwise varying perturbation dynamics. Remarkably, only this additional feedback is required to capture the dynamics of the SSP, which both maintains the turbulent state and enforces its statistical equilibrium.

\section{Conclusions and directions for future work}

In this work we have demonstrated that the RNL system, which models the dynamics of S3T, self-sustains turbulent activity. Comparisons between RNL simulations and the associated DNS demonstrate good agreement between the mean velocity fields and $uv$ component of the Reynolds stress with quantitative differences being confined to the $uu$ Reynolds stress component. The SSP supported by the RNL system is consistent with the familiar roll streak SSP observed in wall turbulence. The results discussed herein suggest that the SSP maintaining turbulence requires both the influence of the perturbations on the streamwise mean flow (captured in the 2D/3C mean flow model) and feedback from the mean flow to the streamwise varying perturbation field, which is additionally retained in the RNL model. Given that the RNL system restricts nonlinear coupling to that between the streamwise mean flow components and the perturbations, this agreement indicates that this highly restricted dynamics captures the fundamental mechanism of turbulence in plane Couette flow.

The RNL model shares the dynamical restriction of the S3T system and is obtained directly from the DNS by eliminating the perturbation-perturbation nonlinearity while retaining the mean-perturbation nonlinearities. It can therefore be seen as providing a bridge between S3T and DNS in which analytic insights gained using S3T can be directly related to DNS, which allows the mechanisms operating in these systems  to be comprehensively compared.
The structure of S3T leads to identification of an analytical SSP. This SSP is associated with the mechanism of streamwise streak forcing by roll circulations which are in turn maintained by perturbation Reynolds stresses~\cite{Farrell-Ioannou-2012}.  Maintenance of the perturbation field in S3T has been shown to result from parametric non-normal interaction between the time-dependent streak and the perturbation field, see e.g. \cite{Farrell-Ioannou-1996b,Farrell-Ioannou-1999, Farrell-Ioannou-2012}.  Construction of the RNL system allows identification of this parametric non-normal SSP, which had been previously demonstrated to be operating in the S3T system, to be extended to DNS turbulence.  The similarity between the SSP operating in the S3T/RNL system and that in DNS provides compelling evidence that the same mechanisms are operating in these systems. Continuing study of these models promises additional insight into the dynamics of turbulence in wall-bounded shear flows.

\iffalse

while at the same time being directly related to the .\fi

%===================================================================
\subsection*{Acknowledgments}
%===================================================================

This work was initiated during the 2012 Center for Turbulence Research Summer Program with financial support from Stanford University and NASA Ames Research Center. We would like to thank Prof. \ P.\ Moin and Prof.\ S.\ Lele for useful comments and fruitful discussions. Financial support from the National Science Foundation under CAREER Award CMMI-06-44793 (to M.R.J. and B.K.L.) and NSF AGS-1246929 and ATM-0736022 (to B.F.F) is gratefully acknowledged.

\section*{Appendix A}
\iffalse
Equation \eqref{eqn:RNL-perturb} can be expressed symbolically as:
\begin{equation}
\mathbf{u}_t =  A(\mathbf{U})  \mathbf{u}+\mathbf{e}~,
\label{eqn:Au}
 \end{equation}
where
$A(\mathbf{U})$ is the linear operator that governs  evolution of the perturbation velocities linearized about the instantaneous $\mathbf{U}(t)$ and ${A} (\mathbf{U})\mathbf{u}$ is a bilinear function of the time-varying streamwise mean flow,
$\mathbf{U}(t)$, and the perturbation velocity, $\mathbf{u}$.  The operator
\fi

The operator $A(\mathbf{U})$ in \eqref{eqn:Au} is obtained by taking the divergence of
\eqref{eqn:RNL-perturb}
 and using continuity \eqref{eq:1c} and $\nabla\cdot \mathbf{e}=0$ to  express the pressure as:
\begin{equation}
p= - \Delta^{-1} \left [  \nabla \cdot \left (   \mathbf{U} \cdot \nabla \mathbf{u} +
\mathbf{u} \cdot \nabla \mathbf{ U} \right ) \right ]~,
\label{eqn:press}
\end{equation}
so that
\begin{equation}
{A} (\mathbf{U})\mathbf{u}= -\mathbf{U} \cdot \nabla \mathbf{u} -
\mathbf{u} \cdot \nabla \mathbf{ U}  + \nabla  \Delta^{-1} \left [  \nabla \cdot \left (   \mathbf{U} \cdot \nabla \mathbf{u} +
\mathbf{u} \cdot \nabla \mathbf{ U} \right ) \right ]  +  \frac{1}{R} \Delta \mathbf{u}~.
\end{equation}
In the above,  $\Delta^{-1}$ is the inverse of  the Laplacian,  rendered unique by imposition of the no slip
boundary conditions at the channel
walls.

%
%
%
%The expression \eqref{eqn:Au} is derived from \eqref{eqn:RNL-perturb} by writing the perturbation in terms of the normal velocity $v$ and normal vorticity $\eta = \partial_z u - \partial_x w$, see e.g.~\cite{Benney-Gustavsson-1981,Schmid-Henningson-2001} eliminating the pressure and then transforming the system back into the $(u,v,w)$ variables as in \citet{Jovanovic-Bamieh-2005}.
%
As described in section \ref{sec:framework}, the S3T system is a second order closure of the NS in \eqref{eq:NSE0}, in which the first order cumulant is  $\mathbf{U}$ and the second order nine component cumulant is  the
spatial covariance  at time $t$ of the flow
velocities $C \equiv C(1,2) = \langle \langle { \mathbf{u}_1 \otimes \mathbf{u}_2 }\rangle  \rangle$ between
the two points $\mathbf{x}_1=(x_1,\,y_1,\,z_1)$ and  $\mathbf{x}_2=(x_2,\,y_2,\,z_2)$
where $\otimes$ is the tensor (outer) product~\cite{Frisch-1995}.
The ensemble average over forcing realizations is denoted by $\langle \langle \cdot \rangle \rangle $,
which under the ergodic assumption is equivalent to
the streamwise average i.e. $\langle \langle \cdot \rangle \rangle \equiv \langle  \cdot\rangle $. The flow then evolves according to \eqref{eqn:S3T}, which is restated here for clarity:
 \begin{subequations}
 \label{eqn:S3TA}
 \begin{align}
{\mathbf{U}_t} = \mathbf{U} \cdot \nabla \mathbf{U}
 + \nabla \mathbf{P} -  \frac{1}{R}\Delta \mathbf{U}  + \mathcal{L}C \tag{\ref{eqn:mean_flow-S3T}} \\
 {C_t} = \left ( A _1(\mathbf{U})+   A _2(\mathbf{U} ) \right )   C + Q \tag{\ref{eqn:perturb-S3T}}
\end{align}
\end{subequations}
where  $\mathbf{A}_1(\mathbf{U}) C  = \langle { \left (\mathbf{A}_1(\mathbf{U} )\mathbf{u}_1 \right ) \otimes \mathbf{u}_2} \rangle$ indicates the contribution to the time rate of change of the covariance from the action of the operator $A(\mathbf{U})$,
evaluated at point $\mathbf{x}_1$,  on the corresponding  component of $C$,  and a similar relation holds for
 $\mathbf{A}_2(\mathbf{U}) C$. $Q = \langle \langle \mathbf{e}_1 \otimes \mathbf{e}_2 \rangle \rangle$ is the second order covariance of the stochastic excitation under the assumption that the noise is temporally delta correlated. The  mean equation
 \eqref{eqn:mean_flow-S3T}  is forced by the divergence of the perturbation
 Reynolds stresses $-\langle \mathbf{u} \cdot \nabla \mathbf{u} \rangle$
 and this term can be expressed as   a linear function of the covariance, $\mathcal{L}C$.

 If \eqref{eqn:mean_flow-S3T}  is considered to be forced independently by a specified Reynolds stress
 divergence specified symbolically as $\mathcal{L}C$, then
 the mean flow dynamics  \eqref{eqn:mean_flow-S3T} define a forced streamwise constant or 2D/3C model of the flow field~\cite{BobbaThesis,Gayme-etal-2010}. The S3T system \eqref{eqn:S3T} is obtained by closing the dynamics through the coupling of the perturbation covariance evolution equation \eqref{eqn:perturb-S3T} to the streamwise constant 2D/3C model.

% \begin{subequations}
% \label{eqn:S3TA}
% \begin{align}
%{\mathbf{U}_t} = \mathbf{U} \cdot \nabla \mathbf{U}
% + \nabla \mathbf{P} -  \frac{1}{R}\Delta \mathbf{U}  + \mathcal{L}C \label{eqn:mean_flow-S3TA} \\
% {C_t} = \left ( A _1(\mathbf{U})+   A _2(\mathbf{U} ) \right )   C + Q \label{eqn:perturb-S3TA}
%\end{align}
%\end{subequations}

\bibliography{S3T_turb_basic_refs}

\end{document}